\documentclass[aps,showpacs,twocolumn,floats,prd,superscriptaddress,nofootinbib]{revtex4} 

\usepackage{amsmath}
\usepackage{latexsym}
\usepackage{graphicx}
\usepackage{epsfig}
\usepackage{verbatim}
\usepackage{hyperref}
\usepackage{url}
\usepackage{color}

\newcommand{\beq}{\begin{equation}}
\newcommand{\eeq}{\end{equation}}
\newcommand{\bea}{\begin{align}}
\newcommand{\eea}{\end{align}}
\newcommand{\bseq}{\begin{subequations}}
\newcommand{\eseq}{\end{subequations}}

\def\leq{\raise 0.4ex\hbox{$<$}\kern -0.8em\lower 0.62ex\hbox{$-$}}
\def\geq{\raise 0.4ex\hbox{$>$}\kern -0.7em\lower 0.62ex\hbox{$-$}}
\def\lsim{\raise 0.4ex\hbox{$<$}\kern -0.8em\lower 0.62ex\hbox{$\sim$}}
\def\gsim{\raise 0.4ex\hbox{$>$}\kern -0.7em\lower 0.62ex\hbox{$\sim$}}
\def\pm{\,\raise 0.4ex\hbox{$+$}\kern -0.8em\lower 0.62ex\hbox{$-$}\,}
\def\fht{f_{\rm HT}}
\def\hht{h_{\rm c, \, \, est.}}

\begin{document}

\title{Binary Systems as Resonance Detectors for Gravitational Waves}
\author{Lam Hui}
\email{lhui@astro.columbia.edu}
\affiliation{Institute for Strings, Cosmology and Astroparticle Physics (ISCAP), Columbia University, New York, NY 10027}
\affiliation{Department of Physics, Columbia University, New York, NY 10027}
\affiliation{Columbia Astrophysics Laboratory, Columbia University, New York, NY 10027}
\author{Sean T. McWilliams}
\email{stmcwill@princeton.edu}
\affiliation{Institute for Strings, Cosmology and Astroparticle Physics (ISCAP), Columbia University, New York, NY 10027}
\affiliation{Department of Physics, Princeton University, Princeton,
  NJ 08544}
\author{I-Sheng Yang}
\email{isheng.yang@gmail.com}
\affiliation{Institute for Strings, Cosmology and Astroparticle Physics (ISCAP), Columbia University, New York, NY 10027}
\affiliation{IOP and GRAPPA, Universiteit van Amsterdam, Science Park
  904, 1090 GL Amsterdam, Netherlands}
\date{\today}


\begin{abstract}
Gravitational waves at suitable frequencies can resonantly interact with
a binary system, inducing changes to its orbit.
A stochastic gravitational-wave background causes the
orbital elements of the binary
to execute a classic random walk, with the variance
of orbital elements growing with time. The lack of such a random walk
in binaries that have been monitored with high precision over long
time-scales can thus
be used to place an upper bound on the gravitational-wave background.
Using periastron time data from the Hulse-Taylor binary pulsar spanning $\sim 30$ years, we
obtain a bound of $h_{\rm c} < 7.9 \times 10^{-14}$ at $\sim 10^{-4}$
Hz, where $h_{\rm c}$ is the strain amplitude per logarithmic frequency interval.
Our constraint complements those from pulsar timing arrays,
which probe much lower frequencies, and ground-based gravitational-wave observations, which
probe much higher frequencies. Interesting sources in our frequency band,
which overlaps the lower sensitive frequencies of proposed space-based observatories,
include white-dwarf/supermassive black-hole binaries
in the early/late stages of inspiral, 
and TeV scale preheating or phase transitions.
The bound improves as (time span)$^{-2}$ and (sampling
rate)$^{-1/2}$. The Hulse-Taylor constraint can be improved
to $\sim 3.8 \times 10^{-15}$ with a suitable observational
campaign over the next decade. 
Our approach can also be applied to other binaries, including
(with suitable care) the Earth-Moon system, 
to obtain constraints at different frequencies.
The observation of additional binary pulsars with the SKA could reach a sensitivity
of $h_{\rm c} \sim 3 \times 10^{-17}$. 
\end{abstract}


\maketitle

\section{Introduction}
The first direct detection of gravitational waves
(GWs) will be a landmark event.  With the advent of the
Advanced Laser Interferometer Gravitational Wave Observatory (LIGO) 
and Advanced Virgo network of detectors 
in $\sim2015$, and the rapid progress of pulsar timing arrays (PTAs), 
it is likely that the next
few years will see this breakthrough come to pass.  

In this paper, we will investigate an alternative approach to
GW detection, based on precision orbital monitoring of binary
systems. The most promising binary systems are those with (at least) one pulsar
member, although the way our method works and the frequencies that are probed differ
significantly from the PTA method. PTAs probe gravitational waves
that pass between the pulsars and us, distorting the arrival times of what
would otherwise be very regular pulses. We are interested instead in
how background GWs interact with the 
orbital dynamics of a binary system. We emphasize
that we are not so much interested in the emission of GWs by the binary -- an important subject in its own right -- 
as in the changes to its orbital parameters due
to scattering with some external GW background.

The scattering is especially effective at GW
frequencies that match a harmonic of the binary's orbital frequency, thereby inducing a resonance.
For a circular orbit, resonance occurs at twice the orbital frequency.
For eccentric orbits, resonance takes place to varying degrees at all harmonics of
the orbital frequency, starting from the fundamental.
Note that this scattering process is quite special and not frequently discussed.  A binary is inevitably losing energy over time by
emitting gravitational waves; the external
GWs merely introduces a small modulation of this overall energy loss
(for external GWs of a sufficiently small amplitude).
Depending on the relative 
phase between the external gravitational wave and the binary, a constructive resonance slows down the energy loss, while a destructive resonance speeds it up.  
Immersed in a stochastic background of GWs, 
the binary orbital elements will thus execute a classic random walk on top of the secular decay due to its own GW emission.  
On average, the expected excursion of orbital elements due to GW scattering vanishes.  However, the variance of the
excursion is non-zero and in fact grows over time, as
in Brownian motion. Such orbital excursions can therefore
be used to detect GWs, and the lack
of such excursions can be used to place bounds.  We again emphasize
the fundamental difference between our approach and that of PTAs,
since we are interested in measuring actual changes in the binary's
orbital parameters, rather than merely apparent changes due to the presence of GWs
along our line of sight.

Versions of this idea have been discussed in pioneering work by
\cite{Rudenko,Mashhoon,Turner,MCH,Bertotti,MTW}.
The subject lay dormant for many years, perhaps due in part to the focus
shifting to PTAs in discussions of pulsars as a detection tool.
We wish to revive the discussion by: 
(1) taking advantage of over 30 years of precision monitoring
of the famed Hulse-Taylor binary, recognizing that
the rms orbital excursion due to GW scattering
increases with time, (2) generalizing earlier work by computing
the random walk of a binary orbit with arbitrary eccentricity, due
to scattering by any stochastic GW background, and (3) finding
the minimum variance estimator for the stochastic GW power spectrum,
given periastron time data; a main result is that the
rms fluctuation in periastron time grows as (time span)$^{3/2}$. 
The approach we propose can be thought of as an astronomical
version of Weber's resonance bar \cite{Weber}.

The only data we thoroughly analyze in this paper is from the Hulse-Taylor (HT)
system PSR B1913+16.  It provides a current constraint on the
stochastic GW background that is weaker
than the one from the Doppler tracking of Cassini, and neither constraint is very restrictive.  
However, future observations of similar binary systems are expected to
improve the constraint considerably. Our method can also be applied to
other binary systems which have also been monitored with high precision over long time scales,
such as the Earth-Moon 
and Earth-Sun systems.
In Sec. V we 
discuss the prospects and challenges of obtaining constraints from them.  We will use the term ``detector binary'' 
as the generic descriptor for the systems of interest.

\section{Formalism} 
We limit our focus to a GW
background that is stochastic in nature, i.e.
during the course of observation, the GW signal is
not dominated by a single source with a definite phase,
but rather arises from a multitude of sources, contributing
to a signal that is statistically stationary.
For practical purposes, this means a Gaussian random field,
although our calculation does not rely on Gaussianity.
Because of our interest in PSR B1913+16, we are particularly
interested in harmonics of its orbital frequency 
$\fht \equiv 3.6 \times 10^{-5}$ Hz. 

This frequency corresponds to
the Hubble scale at a temperature of about $100$ GeV.
Early universe processes, such as preheating after low scale
inflation or bubble collisions at the electroweak phase transition,
generate a stochastic GW background at these frequencies
\cite{Kamionkowski1993,EastherLim2006,Dufaux2007,GarciaBellido2007}.
However, more promising sources of GWs at these frequencies may come
from the later universe, in particular from a large population of
double white dwarf binaries in the early stages of inspiral,
and from supermassive black-hole binaries in the late stages.
These sources emit GWs at frequencies
$f_r = 1.3 \times 10^{-4} {\,\rm Hz} \, [10^5 \, GM c^{-2}/a]^{3/2} \, [M_\odot /M]$,
where $f_r$ is the frequency in the source's rest frame,
$a$ is the mean orbital separation, and $M$ is
the total mass of the binary.
Estimates by \cite{Phinney2001,FarmerPhinney2003,Barack,Timpano,Nissanke} 
suggest that the rms strain amplitude per logarithmic frequency
from white dwarf binaries
is roughly $h_c \sim 10^{-20}$ -- $10^{-19} (f/\fht)^{-2/3}$,
where $f$ is the
orbital frequency of interest, 
and the number of sources within the frequency width
of interest (see below) is
sufficiently large to give a GW background that
is Gaussian random to good approximation.\footnote{Gaussian randomness can be checked, for instance,
by comparing the connected 4th moment against the
second moment squared. Demanding the former is
small compared to the latter is equivalent to requiring
${\rm no. \,\, of \,\, sources}\,\, \times \langle A^2 \rangle^2
/\langle A^4 \rangle \gg 1$,
where $A$ is the amplitude of the gravitational wave from
a given source. For our application,
the number of sources $\sim 10^{11}$ while
$\langle A^2 \rangle^2/\langle A^4 \rangle \sim 10^{-5}$. 
}

Supermassive black-hole binaries constitute another
promising source of GWs. However, the GWs generated by 
these systems at frequencies $\sim 10^{-4}$ Hz would not be stochastic
in character, since the
number of black-hole binaries potentially resonating with HT is much smaller. 
The interaction of the detector binary with GWs
from these sources is therefore better characterized by individual events.
In the course of each event, the relative phase
between the source and the detector binary remains 
coherent; the detector orbital
elements would therefore change in a secular rather than a stochastic
fashion. Over time, the detector binary would
encounter different, uncorrelated events,
and thus there would still be a net random walk of sorts, if viewed over sufficiently long timescales.
However, obtaining quantitative constraints on GWs
of such a character would require a different calculation
from the one presented here -- a subject we hope to
address in the future.
In this paper, we focus instead on the classic random
walk effect, relevant for GWs from double white-dwarf binaries
or the early universe.

Let $(X,Y,Z)$ define the frame of the detector binary, with
the binary orbit lying in the $X$-$Y$ plane.
Let $(x,y,z)$ define the frame of a particular gravitational
wave train, with $\hat{z}$ being
the incident direction. We can go from $(X,Y,Z)$ to $(x,y,z)$ by
performing three consecutive $Z-Y-Z$ Euler rotations.
The last Euler rotation can be ignored since it is equivalent
to rotating among the polarizations of the GWs, which we
average over in any case. Thus, without loss of generality:
\bseq
\label{eq:trans}
\begin{align}
x &= (X\cos\phi + Y\sin\phi)\cos\theta~, \\
y &= Y\cos\phi - X\sin\phi~, \\
z &= (X\cos\phi + Y\sin\phi)\sin\theta~ \, ,
\end{align}
\eseq
where $\phi$ and $\theta$ are two Euler angles.

The induced relative acceleration $\vec{A}\equiv A_x\hat{x}+A_y\hat{y}+A_z\hat{z}$ of a binary system due to a gravitational wave incident
from the $\hat{z}$ direction is given by \cite{MTW}
\bseq
\label{eq:acc}
\begin{align}
A_x &= - R_{x0x0} x - R_{x0y0} y = {1\over 2}
\left( \ddot{h}_{+} x + \ddot{h}_{\times} y \right)~, \\
A_y &= - R_{y0x0} x - R_{y0y0} y = {1\over 2} 
\left( \ddot{h}_{\times} x - \ddot{h}_{+} y \right)~, \\
A_z &= 0\,,
\end{align}
\eseq
where $R_{\mu\nu\alpha\beta}$ is the Riemann tensor, and 
$h$ as the amplitude of the gravitational wave strain,
with $+$ and $\times$ subscripts denoting the two polarization states.
Eq.~\eqref{eq:acc} allows us to write down the time dependence of the
energy,
\begin{eqnarray}
\label{eq:ELdots}
&& \frac{dE}{dt} =\mu\left(A_x\dot{x}+A_y\dot{y}+A_z\dot{z}\right) \nonumber \\
&& \quad \quad ={\mu\over 2}\left(\ddot{h}_{+}(x\dot{x}-y\dot{y})
+\ddot{h}_{\times}(x\dot{y}+y\dot{x})\right)~, 
\end{eqnarray}
where $\mu\equiv\frac{m_1 m_2}{m_1+m_2}$ is the reduced mass of the
binary.  Note that this gives the energy change solely induced by the incoming gravitational wave, which is on top of its original energy loss due to emission.  Analogous expressions can be written down for the angular and center-of-mass linear momenta (see Appendix).

In Eq.~\eqref{eq:ELdots}, we can see that the quantities related to the binary motion are the quadrupole components.  It is straightforward to obtain their Fourier expansions as 
\bseq
\label{eq:transQ}
\begin{align}
X(t)^2 &= X_{\rm o}^2 + \sum_{n=1}^{\infty} Q_{XX}(n) \cos(2\pi n f t)~,\\
Y(t)^2 &= Y_{\rm o}^2 + \sum_{n=1}^{\infty} Q_{YY}(n) \cos(2\pi n f t)~,\\
X(t)Y(t)  &= X_{\rm o}Y_{\rm o} + \sum_{n=1}^{\infty} Q_{XY}(n) \sin(2\pi n f t)~,
\end{align}
\eseq
where $X_{\rm o}$ and $Y_{\rm o}$ are constants, and
$f$ is the orbital frequency.
The $Q$'s are the quadrupole moments \cite{PeterMath}:
\bseq
\label{eq:quadPM}
\begin{align}
Q_{XX}(n) &= \frac{a^2}{n}\bigg(
J_{n-2}(ne)-2eJ_{n-1}(ne) \nonumber \\
&+ 2eJ_{n+1}(ne)-J_{n+2}(ne)\bigg)~, \\
Q_{YY}(n) &= -Q_{XX}(n)+\frac{4a^2}{n^2}J_n(ne)~, \\
Q_{XY}(n) &= \frac{a^2}{n}\sqrt{1-e^2}\bigg(
J_{n-2}(ne)-2J_n(ne)+J_{n+2}(ne)\bigg)~ \, ,
\end{align}
\eseq
where $a$ is the semi-major axis, and $e$ is the eccentricity
($a = 1.95 \times 10^{6}$ km, $e = 0.617$ for PSR B1913+16).
For our purpose, it is convenient to define the following
4 quadrupole moments, which are more closely connected
to the dynamics in the $(x,y,z)$ frame:
\bseq
\label{eq:quads}
\begin{align}
Q_1(n) &= (\cos^2\theta\cos^2\phi-\sin^2\phi)Q_{XX}(n) \nonumber \\
&-(\cos^2\phi-\cos^2\theta\sin^2\phi)Q_{YY}(n)~, \\
Q_2(n) &= \sin2\phi(1+\cos^2\theta)Q_{XY}(n)~, \\
Q_3(n) &= 2\cos\theta\cos2\phi~Q_{XY}(n)~, \\
Q_4(n) &= \cos\theta\sin2\phi\bigg(Q_{XX}(n)-Q_{YY}(n)\bigg)~.
\end{align}
\eseq
For example, Eq.~\eqref{eq:ELdots}
can be rewritten as:
\begin{align}
\frac{dE}{dt} &= \frac{\mu}{4}\sum_{n=1}^{\infty} (2\pi n f) \label{eq-energy} \\
&\bigg(-Q_1(n)~\ddot{h}_{+}~\sin(2\pi n f t)
+ Q_2(n)~\ddot{h}_{+}~\cos(2\pi n f t) \nonumber \\
&+ Q_3(n)~\ddot{h}_{\times}~\cos(2\pi n f t)
+ Q_4(n)~\ddot{h}_{\times}~\sin(2\pi n f t) \bigg)~. \nonumber
\end{align}
Note that $h_{\times}$ and $h_{+}$ have identical statistical properties, 
and are uncorrelated.
Strictly speaking, the orbital motion used on the right hand
side of Eq.~\eqref{eq-energy} to compute $dE/dt$ should be the actual motion,
accounting for both the orbital decay over time due to GW emission,
and the orbital perturbation due to scattering with the external GWs.
However, since both are very small effects, it is a very good
approximation to use the unperturbed orbit.

Depending on the phase of the incoming
strain, $dE/dt$ can take either
sign. Averaging over an ensemble of stochastic GWs
would yield a vanishing change in the orbital energy of a detector binary;
to find an observable signature of GWs, we must therefore compute the energy variance.
Let $\Delta E$ be the energy change 
over some period of time $T$. It can be shown that
its variance takes the form:
\beq
\label{DeltaE2}
\langle \Delta E^2 \rangle = \sum_{i=1}^{4}
\sum_{n=1}^{\infty} \langle \Delta E_i^{(n)} {}^2 \rangle~.
\eeq
where $i$ labels the energy change associated with
the quadrupole $Q_i$. 
As an example, the $i=3$ term is given by
\begin{align}
\langle \Delta E_3^{(n)} {}^2 \rangle &= 
\left({\pi\over 2} n f \mu\right)^2
\int_0^T dt \int_0^T dt'
Q_3(n)^2 \langle \ddot{h}_\times (t) \ddot{h}_\times (t')\rangle \nonumber \\
& \quad \quad
\cos(2\pi n f t) \cos(2\pi n f t')
\label{eq-E3} \nonumber \\
& = T \mu^2 \left(\pi n f\right)^6 {h_c(nf)^2 \over {2nf}} Q_3(n)^2 \, ,
\end{align}
where we use:
\begin{eqnarray}
\langle h_\times (t) h_\times (t') \rangle = 
\langle h_+ (t) h_+ (t') \rangle \nonumber \\ =
{1\over 2} \int_{0}^\infty {df'\over f'} h_c^2(f') e^{i2\pi f' (t - t')} \, ,
\end{eqnarray}
with $h_c^2$ representing the (total) power spectrum
per logarithmic frequency interval. 
We also assume $T \gg 1/(2\pi nf)$,
and use:
\begin{eqnarray}
\label{deltaidentity}
\Big| \int_0^T dt \, e^{i2\pi f' t} {\,\rm cos}(2\pi nft) \Big|^2
\approx {T \over 4} \delta (f' - nf) \, .
\end{eqnarray}
A related delta function identity
explains why different $n$ modes do not mix
in Eq.~\eqref{DeltaE2}. The fact that
different $i$'s do not mix is partly
due to the fact that the two different
polarizations are uncorrelated, and partly
due to the fact that the analog of Eq.~\eqref{deltaidentity}
for mixed $\sin$ and $\cos$ terms vanishes.
Combining all terms, we have:
\begin{align}
\label{eq:delE}
\langle\Delta E^2\rangle &= {1\over 2} \pi^6 T f^5\mu^2 \sum_{n=1}^{\infty}
n^5 h_{\rm c}(nf)^2 \\
&\bigg(Q_1(n)^2 + Q_2(n)^2 + Q_3(n)^2 + Q_4(n)^2\bigg)~. \nonumber
\end{align}
Using the virial relation:
$E = - [G^2(m_1 + m_2)^2 f^2 \pi^2/2]^{1/3} \mu
= - 2\pi^2 f^2 a^2 \mu$, and the period $P=1/f$,
this can be rewritten as:\footnote{
Note that when the system is being perturbed by the external GWs,
the virial relation does not strictly hold on an instantaneous
basis. However, when averaged
over many orbits, we find the virial approximation of relating
changes in energy to changes in period
to be a very good one,
with corrections suppressed by $P/T$;
here the period is defined in an average sense
i.e. $2\pi = \int_T^{T+P(T)} \omega (t) dt$,
as opposed to $2\pi$ divided by the instantaneous angular
velocity $\omega$.
}
\begin{eqnarray}
\label{eq:delP}
{\langle \Delta P^2 \rangle \over P^2}
= {\cal A}^2 \, h_c(2f)^2 {T \over P} \, , 
\end{eqnarray}
with
\begin{eqnarray}
\label{calAdef}
&& {\cal A}^2 \equiv {9\pi^2 \over 32} \sum_n n^5
{h_c(nf)^2 \over h_c(2f)^2} \times \nonumber \\
&& \quad a^{-4} 
\bigg(Q_1(n)^2 + Q_2(n)^2 + Q_3(n)^2 + Q_4(n)^2\bigg) .
\end{eqnarray}
The $Q_i$'s depend on the incidence direction of the GWs,
so we average over $(\theta,\phi)$ to find the net effect.
However, it is worth noting that, even without averaging,
$\langle \Delta E^2 \rangle$
or $\langle \Delta P^2 \rangle$
vary by at most a factor of $2$
across the sky.

Eqs.~\eqref{eq:delE} and \eqref{eq:delP} 
make clear that only harmonics of
the orbital frequency $f$ contribute to the rms energy/period
change -- the hallmark of a resonance effect.
The singling out of these frequencies stems
from delta functions like the one in Eq.~\eqref{deltaidentity},
which has a width $\Delta f \sim 1/T$, where
$T$ is the duration of integration. We are interested
in $T$ from weeks to years ($\gg$ the orbital period of
$0.323$ days for PSR B1913+16), corresponding to 
$\Delta f \sim 10^{-9} - 10^{-6}$ Hz.
GWs within this width of the harmonics would contribute
to the random walk of the binary elements.\footnote{The fact that the cumulative change
in energy $\Delta E$ fluctuates, or random walks,
can be understood as follows.
As $T$ varies, so does $\Delta f$, which controls
which sources of gravitational waves contribute
to resonant scattering with the detector binary.
Since the sources have uncorrelated phases,
$\Delta E$ undergoes random kicks
as the relevant source population varies.
}

If the binary orbit is circular, only
the $n=2$ harmonic contributes, whereas for an eccentric orbit, all 
harmonics including $n=1$ contribute in principle.
In practice, the quadrupole moments $Q_i(n)$ decrease
with $n$, and the expected rms strain $h_c$ drops with frequency,
which counteracts the strong $n^5$ dependence in 
Eqs.~\eqref{eq:delE} and \eqref{eq:delP}.
Assuming $h_c \propto {\rm freq.}^{-2/3}$ and the orbital
parameters of PSR B1913+16, the dominant contributions
come from $n=\{1,4,5,3,6,7,8,2\}$,
in order of importance, with the $n=1$ mode contributing
nearly as much as the other modes combined.
Under the same assumptions, the dimensionless amplitude
${\cal A} \sim 10$. 
(Changing the spectrum to $h_c \propto {\rm freq.}^{-1}$ would only change
${\cal A}$ to $\sim 10.25$.)

Eq.~\eqref{eq:delP} thus tells us that
\begin{eqnarray}
\label{DPrms}
{\Delta P_{\rm rms}
\over P} \sim 10 \, h_c(2f) \sqrt{{T \over P}} \, ,
\end{eqnarray}
which can be understood intuitively as follows.
During each orbital period, the fractional change in period
is roughly given by the strain $h_c$. The cumulative
rms change scales up by the square root of the number
of periods $\sqrt{T/P}$, as expected in a Brownian random walk.
The extra factor of $10$ depends on the 
details: the shape of the orbit and the spectrum of $h_c$ -- we choose
to normalize at $2f$ (twice the orbital frequency) to facilitate
comparison between different binaries, including circular ones.

\section{Data Analysis} 
As a specific application of the generic method derived
in the preceding section, we use
PSR B1913+16 as the detector binary.
The pulsar data carry a wealth of information
about the system. For simplicity, we focus on
the periastron time, recognizing that stronger constraints
could potentially be obtained by analyzing the full time-of-arrival data,
which we leave for future work.
The periastron time data of PSR B1913+16
were published in
\cite{Weisberg}.
They consist of 27 periastron time measurements,
spanning from 1974 to 2006, each of which was obtained from
monitoring the system over approximately 2 weeks (and
$\sim 2$ hours per day over those 2 weeks).
Let us label these times $T_i$, with $i$ ranging from $1$ up to $N$
($N = 27$ in our particular case).
They can be modeled as follows:
\begin{eqnarray}
T_i = \bar T_i + \Delta T_i + n_i \, ,
\end{eqnarray}
where $\bar T_i$ is a smooth component,
$\Delta T_i$ is the excursion induced
by GW scattering, and $n_i$ represents noise.
The smooth component takes the following form:
\begin{eqnarray}
\label{Tbar}
\bar T_i = \alpha + \beta \, p(i) + \gamma \, p(i)^2 \, ,
\end{eqnarray}
where $p(i)$ tracks the number of periods, known
to high accuracy;
$\alpha$ is the zero-point (we choose $p(1)=0$);
$\beta$ is essentially
the period and would be exactly the period 
if there was not a small change over time;
$\gamma$ quantifies the periastron shift
due to the small change in 
the apparent period, induced by
two smooth processes: (1) the famous decay
of the orbit due to the {\it emission} of GWs, and (2)
galactic acceleration of the system as a whole.
Process (1) dominates over (2), though for our purpose
there is no need to differentiate between them.
Both are small compared to the zeroth-order
effect i.e. $\gamma \, p(i) /\beta < 9 \times 10^{-8}$, and we will refer
to $\beta$ as the unperturbed period $\bar P$;
$\gamma$ can be thought of as $\bar P \dot {\bar P}/2$
where $\dot {\bar P}$ is the rate of change of the period
due to (1) and (2).
\footnote{In most of the paper, we simply use $P$
to denote the unperturbed period, when
there is no danger of confusion.}

The fluctuations due to noise $n_i$ and due to
GW scattering $\Delta T_i$ are uncorrelated,
and their respective correlation matrices are
\begin{eqnarray}
\label{CijFij}
\langle n_i n_j \rangle = C_{ij} \quad , \quad
\langle \Delta T_i \Delta T_j \rangle = h_c^2(2\fht) F_{ij} \, ,
\end{eqnarray}
where $C_{ij}$ is the noise matrix, which we treat as diagonal
using error bars from the data, and $F_{ij}$ is defined as
\begin{eqnarray}
\label{Fijdef}
&& F_{ij} \equiv {1\over 6} {\cal A}^2 \bar P {}^2 \left( {\,\rm min}[p(i),
  p(j)] \right)^2
\nonumber \\
&& \quad ( 3 {\,\rm max}[p(i), p(j)] 
- {\,\rm min}[p(i), p(j)] )\,.
\end{eqnarray}
To derive the expression for
$\langle \Delta T_i \Delta T_j \rangle$,
we use the fact that
$\Delta T_i = \int_{T_1}^{T_i} dt \, \Delta P(t)/\bar P$,
and compute $\langle \Delta P(t) \Delta P(t') \rangle$ using
the same technique we used to compute $\langle \Delta P(t)^2 \rangle$
in Eq.~\eqref{eq:delP}.\footnote{We have implicitly assumed
that the periastron time shift from external GWs is
entirely due to their effect on the orbital period.
In reality, there can be an (apparent) periastron time shift
coming from center-of-mass linear momentum
imparted by GWs, or from fluctuations in the orbital eccentricity.
However, the former is suppressed by further powers of the orbital
velocity, and the latter does not lead to a cumulative effect
on the periastron time.
}
We also take advantage of the useful fact that $p(i) = \int_{T_1}^{T_i} dt / P(t)$,
and the expressions above can be derived by noting
that $\Delta P$, $\Delta T_i$ and $\dot {\bar P}$ are
small quantities.
Henceforth, to avoid clutter, we suppress the $i, j$ indices
and use bold-faced symbols to represent matrices or vectors,
wherever no confusion would arise.

It is worth pointing out that Eqs. (\ref{CijFij}) and (\ref{Fijdef})
imply the rms fluctuation in periastron time, for $i=j$, is:
\begin{eqnarray}
\label{DTrms}
\Delta T_{\rm rms} \sim 6 \, h_c(2f_{\rm HT}) P 
\left( {T \over P} \right)^{3/2} \, ,
\end{eqnarray}
where we have abbreviated $T_i$ as $T$, and
$\bar P$ as $P$.
This result can be roughly thought of as coming from
scaling Eq. (\ref{DPrms}) up by $T$. In other words,
the rms fractional change in period {\it per period} is roughly the
strain $h_c$; after a number of periods given by $T/P$,
the rms fractional change in period becomes
$\sim h_c \times \sqrt{T/P}$; 
since the periastron time is cumulatively
dependent on the period, the rms change in periastron time
is $\sim h_c \times \sqrt{T/P} \times T$. 
It is this rapid growth of $\Delta T_{\rm rms}$ with
time span $T$ that we exploit to obtain constraints on $h_c$.

The minimum variance estimator for $\bar {\bf T}$ is
\cite{RybickiPress}:
\begin{eqnarray}
{\bar {\bf T}} {}_{\rm est.} = {\bf L} ({\bf L}^T {\bf C}^{-1} {\bf L})^{-1} {\bf L}^T
{\bf C}^{-1} \, {\bf T}
\end{eqnarray}
where 
${\bf L}$ is an $N \times 3$ matrix:
\begin{eqnarray}
\label{Ldef}
{\bf L} \equiv \left[
\begin{array}{ccc}
1 & p(1) & p(1)^2 \\
1 & p(2) & p(2)^2 \\
:  & : & : \\
1 & p(N) & p(N)^2
\end{array}
\right] \, .
\end{eqnarray}
The corresponding minimum variance estimator for 
the strain power spectrum $h_c^2$ at $2f$ is
\begin{eqnarray}
\hht^2 = \eta \, 
\left( {\bf T}^T {\bf W}^T {\bf C}^{-1} {\bf F}
{\bf C}^{-1} {\bf W} {\bf T} - \Delta \right)
\, ,
\end{eqnarray}
where the matrix ${\bf W}$ is an $N \times N$ matrix defined by
\begin{eqnarray}
{\bf W} \equiv {\bf 1} - {\bf L} ({\bf L}^T {\bf C}^{-1} {\bf L})^{-1} {\bf L}^T
{\bf C}^{-1} \, ,
\end{eqnarray}
and $\eta$ and $\Delta$ are numbers defined by:
\begin{eqnarray}
\eta \equiv 1/{\,\rm Tr.} \left[ {\bf C}^{-1} {\bf F}
{\bf C}^{-1} {\bf W} {\bf F}^T {\bf W}^T \right] \, ,
\end{eqnarray}
\begin{eqnarray}
\Delta \equiv {\rm Tr.} [{\bf C}^{-1} {\bf F} {\bf C}^{-1} {\bf W} {\bf C} {\bf
  W}^T] \, .
\end{eqnarray}
In deriving the above expressions, we have assumed
the variance of the estimators are dominated
by the noise $n_i$ and not the signal $\Delta T_i$.
It can further be shown that the variance
of estimator $\hht^2$ is 
\begin{eqnarray}
\langle \hht^4 \rangle - \langle \hht^2 \rangle^2
= 2\eta
\end{eqnarray}
Because of the subtraction of the
noise power spectrum (the $\Delta$ term), the estimated
$\hht^2$ can be negative, though its
ensemble average cannot.
Note also the estimate $\hht^2$ implicitly assumes
the shape of the power spectrum, through the quantity
${\cal A}$ (see Eq.~\eqref{calAdef}) in the definition of $F_{ij}$. 
We will quote results assuming $h_c (f) \propto f^{-2/3}$.
Assuming the power to go as $f^{-1}$ would alter our results by
a negligible amount compared to the uncertainties involved.

\section{Results}
From the periastron time
data of PSR B1913+16 \cite{Weisberg}, 
we find the power per logarithmic frequency interval
at $2\fht = 7.2 \times 10^{-5}$ Hz to be
$\hht^2 = -6 \pm 7.4 \times 10^{-27}$. 
This is derived from fluctuations in the data
after fitting and removing a smooth quadratic 
(see Eq.~\eqref{Tbar}).  From this, we derive a $95 \%$ upper limit of
$h_c < 7.9 \times 10^{-14}$ at $7.2 \times 10^{-5}$ Hz.\footnote{This upper limit corresponds to
the value of $h_c$ such that the probability
of observing $\hht^2$ at $-6 \times 10^{-27}$ or less
is $5 \%$.}

Pulsars are known to have glitches, which are recognizable by 
abrupt changes in the spin period. The periastron timing
data exhibit a large excursion around the glitch of
May 2003, which we have removed from the above analysis.
Had we included that data point, the results would not
have changed significantly, as we would instead find
$\hht^2 = -3 \pm 7.4 \times 10^{-27}$.

\section{Discussion}

Our GW constraint from the random walk of binary orbital elements
raises several interesting issues.
First, the constraint is a conservative one; namely,
it is based upon a bound on fluctuations of the periastron
data around a smooth curve. If there are additional
sources of fluctuations other than scattering from
the GW background, accounting for them would only
strengthen our bound. However, a more thorough understanding
of the possible sources of fluctuations would be necessary
if one were to claim a detection of the GW background.
Possibilities include glitches and tidal effects.
Glitches are distinguished by accompanying fluctuations
in the pulsar spin period, which are unlikely to have
been caused by stochastic GWs.  Tides cause secular changes
of the binary dynamics rather than stochastic changes. 
To isolate the GW signal from other random or near-random processes,
one can also take advantage of the well defined shape
predicted for the two-point function of the periastron
time fluctuations (see Eq.~\eqref{CijFij}). 

The only other direct bound on GWs at a frequency
$\sim 10^{-4}$ Hz comes from Doppler tracking
of the {\it Cassini} spacecraft \cite{Armstrong}, which is roughly
an order of magnitude more stringent than our constraint.\footnote{
In our method, most of the constraining power of the data
comes from the data points that are furthest apart i.e. $T \sim 32$
years, which corresponds to a fairly narrow frequency window
of $\Delta f \sim 10^{-9}$ Hz.
If one were to extrapolate this amplitude of $h_c$ ($7.9 \times
10^{-14}$) to
a smooth spectrum over a broad bandwidth, one would obtain
$\Omega_{GW} > 1$ which we know is ruled out by
cosmological observations already. From this point of view,
our bound is certainly weak.
But it should be kept in mind that our bound on $h_c$ applies strictly
within a narrow frequency window, for which there is no
useful cosmological bound.
Note also that our bound is weak compared to the bound from big bang
nucleosynthesis (BBN).  
However, once again, the BBN bound assumes a broad (scale invariant)
spectrum of GWs, and it bounds GWs in the early universe and not
from late time astrophysical sources.}
Our bound is rather weak, especially when
compared to the expected GW background from white-dwarf binaries,
as shown in Fig. 1. The expected GW background is
taken from \cite{Barack}, which is consistent with estimates such
as \cite{Timpano, Nissanke}, though with
a large uncertainty.
Also shown is the expected sensitivity
for the proposed eLISA/NGO/SGO detector \cite{LISA}. 

Thus an important question is: how much do we
expect the GW bound to improve
from future observations of PSR B1913+16
and other binaries? To guide our thinking, we observe that the sensitivity
of our method to the GW background scales as
\begin{eqnarray}
h_c &\sim& 5
\left({\delta T \over P}\right)  N^{-{1/2}} \left({T_{\rm tot.} \over
    P}\right)^{-3/2}  \\ \nonumber
		&\sim& 5
\left({\delta T \over P}\right)  n_P^{{1/2}} \left({T_{\rm tot.} \over
    P}\right)^{-2}~,
\end{eqnarray}
where $\delta T$ is the accuracy of each periastron time
measurement, $N$ is the number of such data points, 
$P$ is the orbital period, $T_{\rm tot.}$ is the total
time span, and $n_P$ is the number of periods between 
consecutive periastron time measurements, 
so that $n_P^{-1}$ is the sampling rate.

The minimalist approach would be to simply lengthen $T_{\rm tot.}$.
Assuming the same rate of sampling as before
(one periastron time data point per year), out to the year 2022,
would push the sensitivity on $h_c$ to $8.9 \times 10^{-15}$,
comparable to the {\it Cassini} bound.
This assumes $\delta T \sim 3 \times 10^{-8}$ day, which is
about the level of accuracy towards the later years of the
periastron data we analyzed \cite{Weisberg}.

\begin{figure}[htb]
\begin{center}
\includegraphics[width=0.48\textwidth]{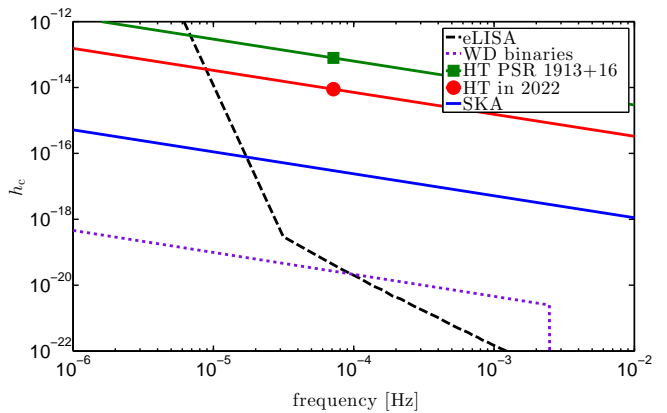}
\end{center}
\caption
{Current constraint and future sensitivity on $h_{\rm c}$, the square
root of the strain power spectrum per logarithmic frequency.
The uppermost line with the (red) square marker shows the current constraint from analyzing
$\sim 30$ years of periastron time data from PSR 1913+16.
The middle line with the (green) circle shows the expected sensitivity from
the same system, if it continues to be observed $\sim2$ hours per day for
$\sim2$ weeks each year until the year 2022.
The bottom (blue) line shows the sensitivity from
a hypothetical observational campaign employing the SKA to observe 100 binary systems (see text for details).
In each case, the solid line going through the dot
represents a $({\rm freq.})^{-2/3}$ spectrum which is assumed in deriving
the bound.
For comparison,
we also show the expected strain sensitivity for eLISA/NGO/SGO \cite{LISA}
(dashed line), and the expected signal strength from white dwarf
binaries
(dotted line) \cite{Barack}. 
We caution that the white dwarf binary background has large uncertainties.
}
\label{fig:hc}
\end{figure}

A more ambitious approach would be to increase the 
sampling as well. Recall that the data we analyzed
came from $\sim 2$ weeks of observations per year during
which the pulsar was observed for only $\sim 2$ hours per day.
This is a fairly sparse sampling.
What if we increase the sampling to one periastron time data point
every 2 weeks (from 2012 to 2022)? The sensitivity on $h_c$
then becomes $3.8 \times 10^{-15}$. 
Increasing the sampling is not as effective
as increasing the total time span, but there is
still some useful gain.

To go beyond this, let us consider the possibilities
offered by the Square Kilometer Array (SKA).
The SKA is expected to find hundreds of binaries with a pulsar member
\cite{smits2009}. Let us assume 100 binaries, 
with an orbital period of around $0.1$ day.\footnote{We envision binaries spanning a range of periods.
Each thus probe the GW background at a different frequency.
Assuming a spectrum for the background, we can
bound a single number, i.e.
the amplitude of the spectrum, using all the binaries.
We assume the external GWs at the 100 binaries can be treated
as uncorrelated. If there is some overlap in frequencies within
the relevant resonant widths, there is the interesting possibility
of cross-correlating excursions between binaries, which we
leave for future work.
}
The SKA also has a higher sensitivity than existing instruments. How much
this translates into an improvement in the pulsar timing residual
depends on how important the pulse jitter is, but an order of
magnitude or so improvement is conceivable \cite{Liu2011}. 
\footnote{In cases where pulse jitter is important, improvement
can only be achieved by longer integration.
}
Let us assume
$\delta T \sim 10^{-9}$ day. (As an example, the periastron time
data had improved in accuracy by almost two orders of magnitude
from 1974 to 2006.) For the sampling, let us use $N = 100$ data points
in the time span of $15$ years. The projected sensitivity
on $h_c$ becomes $\sim 3 \times 10^{-17}$. This is also shown in Fig. 1.

There are a few open questions that remain to be explored.
One is whether an even stronger GW bound can be obtained by analyzing
the time-of-arrival data directly, as opposed to the periastron time
data. There is a wealth of information in the time-of-arrival data,
only a small fraction of which is captured by the periastron time.
The question is whether, as far as the impact of external GWs on
the test binary is concerned, most of the information is already
contained in the periastron time data.
For instance, we have not used any information about changes to
the orbital eccentricity due to scattering with the external GWs, which
can be deduced from the changes in energy
and angular momentum (see Appendix).
How much can our constraint improve if we use such information as
well?
Another interesting question is what current level of constraint we can obtain
from other binary systems. Two in particular come to mind:
the double pulsars PSR J0737-3039A/B, and the Earth-Moon system.
A rough estimate for the double pulsar system, discovered in 2003,
can be obtained by using $\delta T \sim 2 \times 10^{-7}$ day,
$P \sim 0.1$ day, $T_{\rm tot.} \sim 9$ years, $N \sim 30$
\cite{doublepulsar},
giving a current sensitivity to $h_c$ of $\sim 3 \times 10^{-13}$. 
This bound will improve more rapidly than the bound from the HT binary, 
since it has been observed for a shorter time.
The Earth-Moon system is sensitive to the GW background 
at a very different frequency: $\sim 8 \times 10^{-7}$ Hz.
Laser-ranging to the moon can measure
the Earth-Moon distance down to $\sim 15$ mm, 
which is a fractional accuracy of about $4 \times 10^{-11}$
\cite{Murphy12}. The main hurdle to obtaining accurate constraints
on GWs is the need to model many geophysical effects of
both the Earth and the Moon. A conservative bound can be
obtained as long as one does not over-fit the lunar-ranging data.

Let us close by noting that our calculation applies strictly to
a stochastic background. The case of supermassive black hole
mergers needs to be separately considered, since their
relative scarcity implies a GW signal more in the form of
individual events, each of which is coherent.
We hope to explore this in a future paper.

\vspace{0.1in}

\section*{Acknowledgements} 
We thank Adam Brown, Fernando Camilo, Sergei Dubovsky, Eanna Flanagan, 
Eric Gotthelf, Zoltan Haiman, David Hogg, Mike Kesden, Vicky Kaspi,
Michael Kramer, Szabi Marka, Wei-Tou Ni,
Alberto Nicolis, Joel Weisberg and Matias Zaldarriaga for
useful discussions. 
We are grateful to Jim Cordes for helpful discussions, and comments from Marc Kamionkowski that sharpened our explanation of several points.
We especially thank Joe Taylor
for providing the periastron time data and patiently answering
many questions about them.
This work was supported by the DOE, NASA and
NSF under cooperative agreements
DE-FG02-92-ER40699, NNX10AN14G, AST-0908365, and PHY11-25915.
The work of I-Sheng Yang is supported in part by the Foundation for Fundamental Research on Matter (FOM), which is part of the Netherlands Organisation for Scientific Research (NWO).

\appendix

\section{Angular momentum}

For completeness, we provide equations that describe the
change of angular momentum due to scattering with the external GWs.
\begin{eqnarray}
\label{eq:Ldots}
&& \frac{dL_{x}}{dt} =\mu\left(-za_y\right)
={1\over 2} \mu\left(\ddot{h}_{+}yz-\ddot{h}_{\times}xz\right)~, \nonumber \\
&& \frac{dL_{y}}{dt} =\mu\left(za_x\right)
={1\over 2} \mu\left(\ddot{h}_{+}xz+\ddot{h}_{\times}yz\right)~, \nonumber \\
&& \frac{dL_{z}}{dt} =\mu\left(xa_y-ya_x\right) \nonumber \\
&& \quad \quad ={1\over 2}\mu\left(-2\ddot{h}_{+}xy+\ddot{h}_{\times}(x^2-y^2)\right)~,
\end{eqnarray}
Applying the same procedure as we did for the energy,
the variance in the change in angular momentum is:
\bseq
\label{eq:Lcomp}
\begin{align}
\langle(\Delta L_x)^2\rangle &= \langle(\Delta L_y)^2\rangle =
\frac{\pi^4}{8} T f^3\mu^2 \sum_{n=1}^{\infty}
n^3 h_{\rm c}(nf)^2  \\
&\bigg(Q_5(n)^2 + Q_6(n)^2 + Q_7(n)^2 + Q_8(n)^2\bigg)~, 
\nonumber \\
\langle(\Delta L_z)^2\rangle &=
{1\over 2} \pi^4 T f^3\mu^2 \sum_{n=1}^{\infty}
n^3 h_{\rm c}(nf)^2  \\ 
&\bigg(Q_1(n)^2 + Q_2(n)^2 + Q_3(n)^2 + Q_4(n)^2\bigg)~ , \nonumber 
\end{align}
\eseq
where $Q_5, Q_6, Q_7, Q_8$ are defined by
\bseq
\label{eq:quads5}
\begin{align}
Q_5(n) &= \sin2\theta
\bigg(Q_{XX}(n)\cos^2\phi+Q_{YY}(n)\sin^2\phi\bigg)~, \nonumber \\
Q_6(n) &= \sin2\phi\sin2\theta~Q_{XY}(n)~, \\
Q_7(n) &= 2\sin\theta\cos2\phi~Q_{XY}(n)~, \\
Q_8(n) &= \sin\theta\sin2\phi\bigg(Q_{XX}(n)-Q_{YY}(n)\bigg)~. 
\end{align}
\eseq
The variance in the magnitude of ${\vec L}$ is given by
\begin{align}
\label{eq:Ltot}
\langle (\Delta|\vec{L}|)^2 \rangle &=
|\vec{L}|^{-2}\langle(\vec{L}\cdot\vec{\Delta L})^{2}\rangle \nonumber \\
&= |\vec{L}|^{-2}\langle(L_x \Delta L_x +
L_y \Delta L_y + L_z \Delta L_z)^{2}\rangle \nonumber \\
&= |\vec{L}|^{-2}\bigg(
L_x^2\langle (\Delta L_x)^2\rangle+
L_y^2\langle (\Delta L_y)^2\rangle   \nonumber \\
&+L_z^2\langle (\Delta L_z)^2\rangle
\bigg) \nonumber \\
&= \sin^2\theta \langle(\Delta L_x)^2\rangle + \cos^2\theta
\langle (\Delta L_z)^2 \rangle~.
\end{align}

\bibliographystyle{prsty}
\bibliography{references}

\end{document}